# Uncertainty Quantification as a Complementary Latent Health Indicator for Remaining Useful Life Prediction on Turbofan Engines

L. Thil*, J. Read**, R. Kaddah***and G. Doquet****

*LIX Ecole Polytechnique & IRT SystemX: thil@lix.polytechnique.fr
** LIX Ecole Polytechnique: jesse.read@polytechnique.edu
*** IRT SystemX: rim.kaddah@irt-systemx.fr
**** Safran Tech: guillaume.doquet@safrangroup.fr

**Abstract**. Health Indicators (HIs) are essential for predicting system failures in predictive maintenance. While methods like RaPP (Reconstruction along Projected Pathways) improve traditional HI approaches by leveraging autoencoder latent spaces, their performance can be hindered by both aleatoric and epistemic uncertainties. In this paper, we propose a novel framework that integrates uncertainty quantification into autoencoder-based latent spaces, enhancing RaPP-generated HIs. We demonstrate that separating aleatoric uncertainty from epistemic uncertainty and cross combining HI information is the driver of accuracy improvements in Remaining Useful Life (RUL) prediction. Our method employs both standard and variational autoencoders to construct these HIs, which are then used to train a machine learning model for RUL prediction. Benchmarked on the NASA C-MAPSS turbofan dataset, our approach outperforms traditional HI-based methods and end-to-end RUL prediction models and is competitive with RUL estimation methods. These results underscore the importance of uncertainty quantification in health assessment and showcase its significant impact on predictive performance when incorporated into the HI construction process.

## 1. Introduction

Accurate estimation of the RUL of complex systems, such as industrial machinery or aerospace components, is critical for enabling predictive maintenance, reducing operational costs, and mitigating catastrophic failures. By translating raw sensor data into HIs that reflect system degradation, RUL prognostics provide actionable insights for real-time decision-making. Traditional approaches to prognostics have relied on statistical methods, including regression models, survival analysis, and stochastic techniques, which often require strong assumptions about degradation patterns or failure modes. However, the advent of deep learning has allowed the field to leverage data-driven models to autonomously learn complex degradation trends directly from high-dimensional sensor data, eliminating the need for handcrafted feature engineering. Among these data-driven methods, autoencoder (AE)-based frameworks have gained prominence for their ability to learn healthy system representations. By training on nominal operational data, AEs reconstruct input signals, and deviations in reconstruction error serve as a measure of novelty detection indicating a degradation from this nominal healthy state. Recent advances, such as RaPP Ki Hyun et al. (2020) and González-Muñiz et al. (2022), refine this approach by iteratively projecting reconstruction errors into the latent space of the encoder, generating more accurate HIs to model degradation trends. While promising, such methods often overlook the inherent uncertainty in model predictions. This is a critical gap for safety-critical applications where unreliable HIs can led to costly overhauls or unexpected downtime.

In this work, we propose to augment the RaPP HIs with uncertainty quantification (UQ) through Monte Carlo dropout, enabling probabilistic interpretation of degradation signals. Our contributions are in the fusion of these indicators and benchmarking the HIs directly on RUL estimation for prognosability clarity.

## 2. Literature Review

There exist various methods to estimate degradations in systems using data-driven approaches, such as measuring a degradation trend, doing linear regression. But a drawback from these is that often they are not interpretable, or do not offer the possibility to assess in real time the current system's status for a domain expert to comprehend. Therefore, the notion of health indicators (HIs) is crucial to produce a set of quantifiable metrics providing a real time description of the health of the system. Common HIs could be the remaining capacity of a lithium-ion battery, the State of Health (SOH) or any relevant metric that provides an assessment of a system or its inner key components.

The State of Health (SOH) of unmaintained engineering systems is modeled as a continuous monotonous function, reflecting the expectation that such systems degrade progressively until failure without self-recovery or maintenance. Health Indicators (HIs) must therefore provide actionable insights into this degradation trajectory. Common evaluation criteria for HIs include monotonicity (consistent degradation trend), trendability (detectable degradation patterns), and prognosability (predictive utility for remaining useful life, or RUL) (Coble, 2010). Traditional data-driven HIs rely on statistical features of sensor signals, such as skewness, kurtosis, or variance over time, whose evolution correlates with degradation states (Fink et al., 2020). However, emerging approaches leverage machine learning to design more robust HIs. Among these, autoencoders (AEs) have gained prominence: trained on healthy, non-degraded data (e.g., sensor readings from intact turbofan engines), AEs learn to reconstruct normal operating patterns. When exposed to degraded inputs (e.g., signals from aging components), the AE's reconstruction error—measured as the difference between the input (original data) and output (reconstructed data)—increases, serving as a direct HI (Huang et al., 2023). This error, often quantified via metrics like mean squared error (MSE), acts as a novelty score, where higher deviations signal greater degradation severity. For instance, in turbofan engine monitoring, the rising reconstruction error mirrors progressive mechanical wear, enabling RUL estimation. While most methods focus on input-output reconstruction, recent work exploits latent space distances in AEs, leveraging their disentangled representations to capture degradation patterns more robustly.

Current metrics focus on the entire input/output space, and these individual parameters do not necessarily reflect the hierarchical nature of deep learning models in the inner latent spaces. Comes into play the RaPP by Ki Hyun et al. (2020) which uses the projections of the activations of the model's parameters to compute the errors. It is based on two families: one is a Euclidean distance of the activations of the model's parameters computed by each layer, and the second is based on the Mahalanobis distance as it takes into consideration the variance and correlation of the different components. These RaPP metrics were initially proposed in the context of an encoder hidden layer space, but González-Muñiz et al. (2022) highlighted better results when it was used only in the latent space of the AE architecture, further shown in Thil et al. (2025).

In UQ, noise is categorized as aleatoric when arising from inherent data variability that cannot be reduced with additional samples, or epistemic originating from model uncertainty, which diminishes with more training data. Building on this distinction, Akbari and Jafari (2020) leveraged a Variational Auto-Encoder (VAE) with Monte Carlo dropout Gal & Ghahramani (2015) between its latent space and decoder layers to disentangle these uncertainties. While prior work highlights methodological advances, the interplay of HIs for degradation modeling remains understudied. In this study, we investigate how integrating such HIs enhances data-driven degradation modeling, with a focus on their combined efficacy for accurate remaining useful life (RUL) estimation.

## 3. Methods

### 3.1 Reconstruction and Latent Distance Measures

In recent years, autoencoders architectures have been used as HI by using the reconstruction error of its output $\hat{x}$ from an input $x$. The underlying idea is to train an autoencoder architecture over non-degraded inputs such that when presented with degraded inputs, it will fail to perform an accurate reconstruction. This error can be measured to identify degraded samples, while quantifying their degree with a distance measure based on the reconstruction error $\epsilon_{REC}$ defined as:

$$\epsilon_{REC} = |x - \hat{x}|. \quad (1)$$

Current metrics focus on the entire input/output space, and these individual parameters do not necessarily reflect the hierarchical nature of deep learning models in the inner latent spaces Ki Hyun et al. (2020). Using a larger manifold with the hidden representation of an encoder shows superior capabilities in novel detection than measuring only $\epsilon_{REC}$ (1).

The RaPP methods use the projections of the activations of the model's parameters to compute the errors. It is based on two families: $\epsilon_{SAP}$ which is a Euclidean distance of the activations of the model's parameters computed by each layer, and $\epsilon_{NAP}$ which is based on the Mahalanobis distance and and differs from $\epsilon_{SAP}$ as it takes into consideration the variance and correlation of the different parameters. $\epsilon_{NAP}$ attributes less weight to sparser components. In the case of signal analysis with the assumption that each component be a random Gaussian variable, it leads to minimizing the influence of the noisiest components in the computation of the distance metric.

$$\epsilon_{SAP}(x) = |h_i(x) - h_i(\hat{x})|. \tag{2}$$

$$\epsilon_{NAP}(x) = |(d(x) - \mu_X)^T V \Sigma^{-1}|. \tag{3}$$

Where $h_i(x)$ are the activations of the ith hidden layer $h$ from input $x$, $d(x) = h(x) - h(\hat{x})$, $\mu_X$ is the column wise mean of the matrix $D$ with the rows collecting $d(x)$, and $V$ containing the right singular vectors from the SVD of $D$, and $\Sigma$ the diagonal of $\bar{D}$ denoting the column-wise centered version of $D$. These RaPP metrics were initially proposed in the context of an Encoder space, but González-Muñiz et al. (2022) highlighted better results when it was used only in the latent space of the AE architecture. Similarly, we denote the RaPP HIs from the encoder as $\epsilon_{SAP}$ and $\epsilon_{NAP}$, and the latent those computed on $z$ only noted $\epsilon_{SAP_{LS}}$ and $\epsilon_{SAP_{LS}}$.

3.2 *Adding Uncertainty Quantification: Aleatoric versus Epistemic*

Aleatoric uncertainty noted $\sigma_a$ comes from noise in the data and cannot be reduced by adding more data, whereas epistemic uncertainty $\sigma_e$ originates from the lack of knowledge of the model and can be mitigated with more samples. Uncertainty in an autoencoder (AE) model can be quantified using Monte Carlo dropout applied to the decoder. However, since aleatoric noise cannot be explicitly modeled in the latent space of an AE, disentangling aleatoric and epistemic uncertainties remains challenging. In contrast, variational autoencoders (VAEs) offer a more nuanced approach. By modeling the latent space as a Gaussian distribution, VAEs enable the separation of these two types of uncertainties, providing a clearer distinction between inherent data noise and model-related uncertainties.

In Bayesian deep learning, encoder model parameters $W_E$ and latent variables $z$ are treated probabilistically. For regression tasks, the likelihood is modeled by a Gaussian:

$$p(y|W,x) = \mathcal{N}(y; \mu^{W_E}(x), \sigma^{W_E}(x)^2). \tag{4}$$

Where $\mu^{W_E}(x)$ and $\sigma^{W_E}$ are encoder-derived mean and variance. MC dropout approximates Bayesian inference by sampling from the posterior via stochastic dropout mask during inference. The predictive distribution integrates over both W and latent variables $z$:

$$p(y|x) = \int N(y; \mu^{W_E}(x,z), \sigma^{W_E}(x,z)^2) p(W|X) p(z|x) dW_E dz. \tag{5}$$

where X is the training data. This intractable integral is approximated via MC sampling:

$$p(y|x) \approx \frac{1}{n} \sum p(y|\hat{W}_E, x, \hat{z}). \tag{6}$$

With $\hat{W}_E \sim p(W_E)$ (via dropout) and $\hat{z} \sim p(z|x)$. VAE explicitly models aleatoric uncertainty through a stochastic latent space $\hat{z} \sim q(z|x)$ is the encoder's approximate posterior. The training objective combines reconstruction loss and KL divergence regularization:

$$\mathcal{L}_{VAE} = -E_{q(z|x)}[\log p(x|z)] + \beta \cdot D_{KL}(q(z|x)|p(z)). \tag{7}$$

With $p(z)$ as a standard Gaussian prior. During inference: 1) Aleatoric uncertainty is sampled via $\hat{z} \sim q(z|x)$, 2) Epistemic uncertainty is introduced through dropout on decoder weights $W_E$. In the case of vanilla AEs, aleatoric uncertainty $\sigma_a$ cannot be isolated as $z$ is deterministic, rendering it undefined. This highlights the advantage of VAEs for joint uncertainty estimation.

The randomness of the latent space $z$ can be used to quantify the aleatoric uncertainties, and where $W_h$ being the weights of a classification head can be used to quantify the epistemic uncertainties by applying dropout over multiple passes. To perform Monte Carlo sampling, from an input $x$ we infer the model over $n$ passes while applying dropout over the weights and averaging the outputs as the Bayesian inference. Critically, because this is a reconstruction regression task, we perform UQ on the full output rather than a specific label, but we need a reduction operation on the dimensions to have the uncertainty value to 1. As such, UQ is applied to the reconstruction error $\epsilon_{REC}$ instead of the direct full output $y$.

Finally, our AE model will model a global noise that is likely to contain mostly epistemic noise, whereas in the VAE we will be able to separate the two. We will use the uncertainties quantifications over the data as standalone HIs that we can combine with the RaPP ones.

3.3 *Creating and Benchmarking Health Indicators*

Compared to previous HI approaches using RaPP, we will use these series to predict a RUL. We train a random forest (RF) regressor over those and perform an ablation study to assess their impact on predicting a RUL. Furthermore, this benchmark will allow us to directly compare the predictive capabilities of these indicators compared to more traditional models trained directly to predict the RUL from direct sensor data. Because these latter approaches do not produce HIs they do not have clear evaluation metrics over the values we produce. In this work, we bridge this gap by benchmarking our created HIs with existing methods.

There are various benchmarks performed over C-MAPSS Saxena et al. (2008) but they often carry different protocols. Indeed, the underlying degradation mechanism in C-MAPSS is rather linear until it enters an exponential growth in the latter stages. Thus, many approaches defined a cutoff threshold called $R_{early}$ that was commonly used at RUL=125, sometimes they would clip values larger than 125 as they assumed that the degradation was not sufficiently present.

3.4 *Datasets: 4 subsets from the C-MAPSS repository with a mix of flight conditions and degradation mechanisms*

The train trajectories are complete until a point of failure indicated by RUL=0, while the test set cuts off before failure. In some of the protocols over the test set, we only predict up to that point. The datasets contain an increasing number of conditions and fault modes, and it can be assumed that the difficulty increases for each set, FD001 being the simplest one and FD004 the hardest.

|  | FD001 | FD002 | FD003 | FD004 |
|---|---|---|---|---|
| # Train trajectories | 100 | 260 | 100 | 248 |
| # Test trajectories | 100 | 259 | 100 | 249 |
| # Conditions | 1 | 6 | 1 | 6 |
| # Fault modes | 1 | 1 | 2 | 2 |

Table 1. Description of the C-MAPSS Dataset

3.5 *Evaluating the Health Indicators*

Classical HIs are assessed through monotonicity, trendability, and prognosability. However, monotonicity and trendability lack direct relevance to SOH estimation, offering limited interpretability and inconsistent performance insights. Prognosability often yields near-perfect scores (e.g., 0.99), misleadingly implying precise SOH knowledge despite poor RUL prediction accuracy. To overcome these limitations, we propose benchmarking HIs directly on RUL estimation using a RF regressor, grounding evaluation in empirical performance. An ablation study further examines HI combinations' impact on prediction efficacy. Our work evaluates the complementary role of UQ with RaPP-generated HIs in RUL estimation, contrasting results against conventional HI metrics.

## 4. Results

We show that both AE in Fig. 1 and VAE in Fig. 2 models demonstrate consistent reductions in RMSE when combining HIs derived from RaPP with UQ. For instance, on dataset FD001, the combined use of latent RaPP features (NAP and SAP) and UQ achieves near state-of-the-art (SOTA) performance (RMSE = 11.23 vs. SOTA = 11.17). Notably, our approach surpasses existing benchmarks on more complex datasets (FD003 and FD004), where the inclusion of diverse HIs leads to superior RUL estimation accuracy. We attribute this improvement to the complementary behavior of combined HIs: while individual indicators exhibit oscillatory patterns, their integration stabilizes the health status representation, enabling clearer degradation trends.

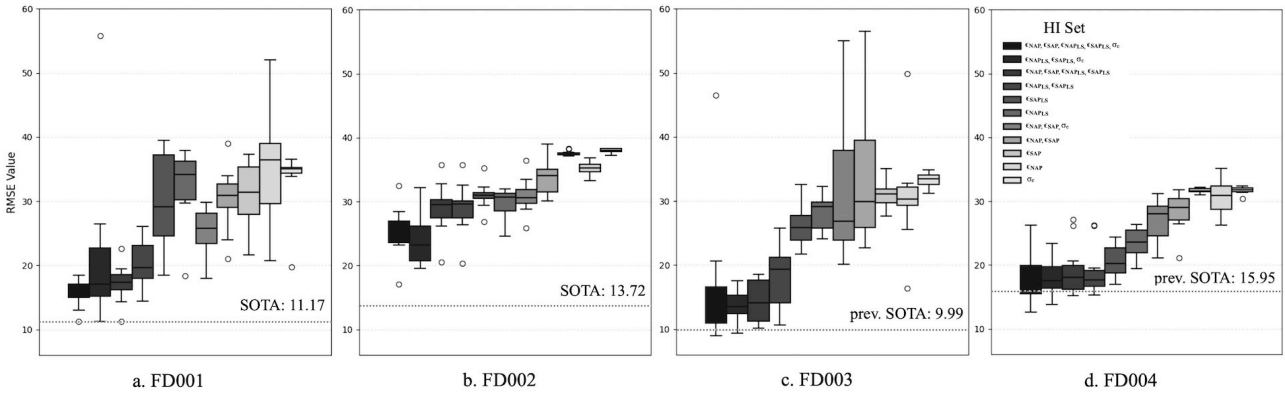

Figure 1. Autoencoder (AE) benchmark results over C-MAPSS dataset for different HI groups. Lower RMSE scores means better accuracy in RUL estimation. We observe that the groups combining the different HIs together achieve SOTA it for FD003 and FD004 thought as the most complex groups despite using a simple RF regressor.

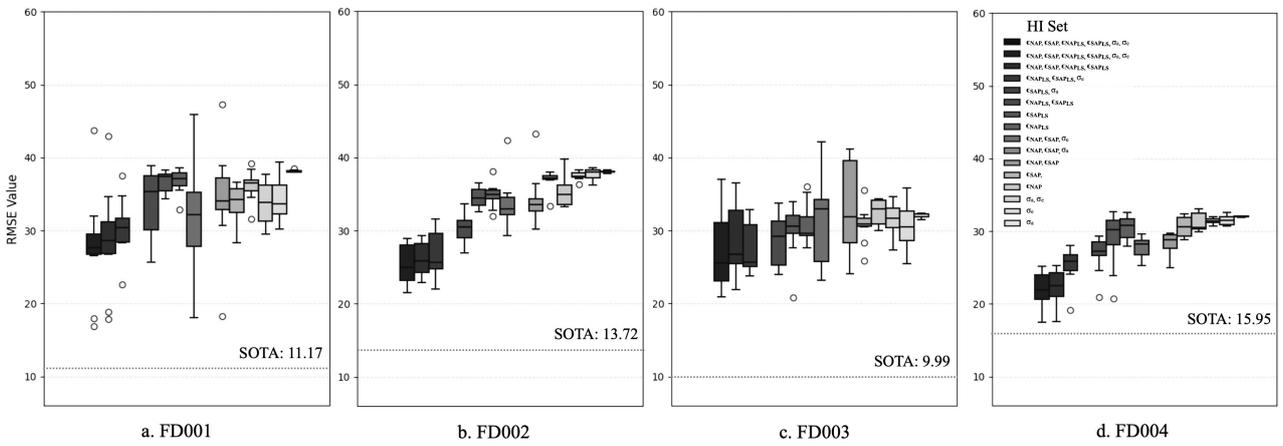

Figure 2. Variational Autoencoder (VAE) benchmark results over C-MAPSS dataset for different HI groups. We do not find better results than the AE counterpart shown in fig. 1 contrary to the expectations that the VAE architecture outperforms it. This is likely explained by the high noise in the data, but on FD004 the results are converging near SOTA performance suggesting potential over more complex datasets.

| HI | FD001 | | | | FD003 | | | |
|---|---|---|---|---|---|---|---|---|
|  | Mono | Trend | Progno | RUL RMSE | Mono | Trend | Progno | RUL RMSE |
| $\sigma_e$ | 0,601 | 5,900E-05 | 0,748 | 19,74 | 0,513 | 4,000E-05 | 0,617 | 31,21 |
| $\epsilon_{SAP}$ | 0,691 | 3,320E-04 | **0,850** | 21,64 | 0,821 | 1,430E-04 | **0,774** | 27,63 |
| $\epsilon_{NAP}$ | 0,330 | 3,700E-05 | 0,617 | 20,67 | 0,421 | 1,400E-05 | 0,148 | **16,4** |
| $\epsilon_{SAP_{LS}}$ | 0,728 | 5,830E-04 | 0,883 | 18,42 | 0,711 | 2,700E-05 | 0,325 | 21,66 |
| $\epsilon_{NAP_{LS}}$ | 0,395 | 3,200E-05 | **0,545** | 18,32 | 0,306 | 1,900E-05 | 0,630 | 24,05 |

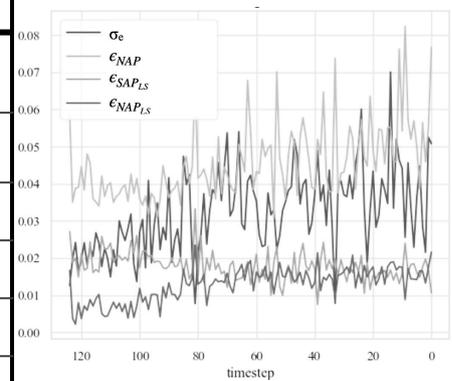

Table 2: Example of HIs evaluation metrics in respect to RUL RMSE (lower is better). The best RUL estimation metrics over C-MAPSS are not necessarily those showing the best prognosability (in bold), or with monotonicity and trendability.

Figure 3: HI Evolution of engine 1 from FD001 test. Encoder_SAP not displayed because it has different scale values.

To further evaluate HI quality, Fig. 3 compares traditional metrics (monotonicity, trendability, prognosability) against RUL prediction RMSE. Results reveal weak links between these classical metrics and actual prognostic performance. For example, the feature achieves the lowest prognosability score but delivers the best RMSE for RUL prediction. This discrepancy highlights the limitations of relying solely on conventional HI evaluation criteria and underscores the need for task-specific metric alignment.

## 5. Discussion and Conclusions

We showed that coupling different HIs computed in the latent space of deep learning models is an effective approach in modeling degradation dynamics. We've shown that adding an uncertainty quantification component made the predictions more robust. To adapt the evaluation of our HIs, we propose to diverge from the traditional metrics of monotonicity, trendability and prognosability because they poorly reflect degradation mechanisms, and we instead propose to directly benchmark on RUL prediction to assess their quality. This also allows for an ablation study between different HI combinations to assess their performance. In short, the RaPP metric combined with uncertainty quantification isolating aleatoric and epistemic noise is a solid approach to model degradation dynamics from HIs in the latent space of deep learning models.

For future works, we aim at testing with architectures combining different window sizes, better regressor to evaluate the HIs or propose an architecture allowing to isolate sub-system HIs. The next area of focus would be over robustness and interpretability of HIs generated from the latent space of deep learning models.

## Acknowledgements

This work has been supported by the French government under the "France 2030" program, as part of the SystemX Technological Research Institute within the JNI3 project.